%% file: main_acs.tex
\documentclass[journal=jctcce,manuscript=article]{achemso}
\usepackage[version=3]{mhchem} 
\usepackage{multirow}
\usepackage{threeparttable}
\usepackage[labelsep=period]{caption}
\usepackage{appendix}
\usepackage{braket}
\usepackage{caption}
\usepackage{subcaption}
\usepackage{comment}
\usepackage{color}
\usepackage{dcolumn}
\usepackage{graphicx}
\usepackage{dcolumn}
\usepackage{bm}
\usepackage[section]{placeins}
\usepackage{soul}
\usepackage{siunitx}
\usepackage[table]{xcolor}
\usepackage{amsmath}
\usepackage{algorithm}
\usepackage{algpseudocode}
\usepackage{nicefrac}

\usepackage{titlesec}
\titleformat{\section}[block]
  {\fontsize{16}{16}\bfseries}
  {\thesection}
  {1em}
  {}
\titleformat{\subsection}[hang]
  {\fontsize{14}{14}\bfseries}
  {\thesubsection}
  {1em}
  {}  

\author{Sonja Bumann}
\affiliation{Department of Chemistry, University of California, Berkeley, California 94720, USA }
\alsoaffiliation{Chemical Sciences Division, Lawrence Berkeley National Laboratory, Berkeley, CA, 94720, USA}

\author{Eric Neuscamman}
\email{eneuscamman@berkeley.edu}
\affiliation{Department of Chemistry, University of California, Berkeley, California 94720, USA }
\alsoaffiliation{Chemical Sciences Division, Lawrence Berkeley National Laboratory, Berkeley, CA, 94720, USA}

\title
  {Reducing the Cost of Energy Differences in Variational Monte Carlo
   with Spotlight Sampling}

\begin{document}

\begin{tocentry}
\begin{center}
\includegraphics{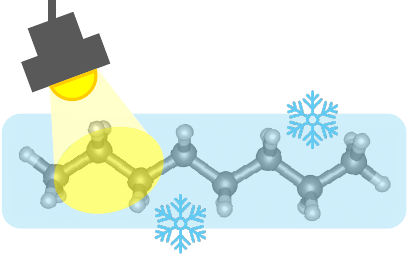}   
\end{center}
\end{tocentry}

\begin{abstract}
We investigate an approximate sampling scheme that can significantly
reduce the cost scaling of variational Monte Carlo when it is
employed to predict the energy differences associated with local
chemical changes.
Inspired by side-chaining and embedding methods, this spotlight
sampling approach adopts an approximate fragmented Hamiltonian
and correlated sampling to reduce cost scaling to the point
that it is essentially linear with system size, with the
potential to go sub-linear if certain conditions are met.
In tests on bond stretching energies in alcohols, hydrogen
dimer chains, and molecules with various degrees of $\pi$-system delocalization, we observe the anticipated linear scaling
as well as an explicit cost crossover with
standard variational Monte Carlo.
\end{abstract}


\input{introduction.tex}
\input{theory.tex}

\input{results.tex}

\input{conclusion.tex}
\input{appendix.tex}

\begin{acknowledgement}

\input{acknowledgement.tex}

\end{acknowledgement}

\FloatBarrier

\nocite{*}
\bibliography{main}

\end{document}

%% file: introduction.tex
\section{Introduction}

Simultaneously capturing the effects of both weak and strong electron correlation
remains challenging for traditional quantum chemistry methods, especially in
excited states and large systems.
\cite{Levine2006, SHERRILL_bond_breaking, DipTM2019,Aoto2017,VVCC2000, DFT_bad_at_TM, CASSCF_limit, Olsen_casscf}
In many difficult cases, Quantum Monte Carlo (QMC) methods have proven to be
effective alternatives. \cite{Shee_AFQMC, Needs_DMC, QMCPack, Clark2011, Assaraf2017, Petruzielo, Giner2016, Dash2019, Dash2021, Sergio2019, LeonRev, Leon2020}
For example, in some excited states,
even variational Monte Carlo (VMC), perhaps the most straightforward QMC technique,
has proven more accurate than full-triples equation of motion coupled cluster theory.
\cite{LeonRev, Leon2020}
However, the significant computational cost required to evaluate large numbers of random samples
continues to limit the applicability of both VMC and QMC methods more generally.
Indeed, such sampling places an unusually large prefactor in front of VMC's
quartic cost growth with system size, \cite{Bienvenu2022}
meaning that the advantages of this modest polynomial scaling can be
difficult to realize in practice. In this study, we show that a combination of approximate fragmentation and correlated
sampling can reduce both cost and scaling while maintaining accuracy
when using VMC to predict energy differences
associated with local chemical changes, as for example when stretching a chemical bond.

It has long been recognized that, thanks to correlated sampling techniques, it can be
advantageous to directly estimate energy differences in VMC rather than attempting
to make independent estimates of the energy before and after a chemical change. \cite{SWCT1}
The basic idea is that, for regions of the system that are only weakly
affected by the change, statistical fluctuations will mostly cancel out so long
as essentially the sampling of electron positions in those regions is used
on either side of the energy difference.
The space warp transformation, which we employ in the present study,
achieves this goal explicitly. \cite{SWCT2, space_warp}
What's more, by trading the task of estimating two extensive energies for
the task of estimating an intensive energy difference, such correlated sampling
approaches should in general drop the cost scaling of VMC from quartic to cubic.
Here, we seek to combine this advantage with the advantages of local fragmentation
approaches, in particular approaches inspired by recent side-chaining methods.
\cite{Bienvenu2022,Feldt2021}

Locality has, more generally, long been exploited in attempts to mitigate the cost of VMC.
\cite{Kohn1959, Maslen1998, LinHF, LinHF2}
Early works focused on leveraging the sparsity of the Slater matrix
when using local orbitals in order to reduce the costs associated with
Slater determinant ratios,
\cite{Williamson2001, ML2003, DAlfe2004, Austin2006}
which can also be mitigated by complimentary delayed update methods.
\cite{mcdaniel2017delayed}
More recently, Feldt and coworkers have shown that that evaluating local side chains,
in which only a few of the electrons near a given nucleus are allowed to
move, is an effective way to reduce the cost of dealing with core electrons.
\cite{Feldt2021}
This side-chaining approach was then generalized in a study by Bienvenu and coworkers,
in which they show that it can reduce the cost of VMC sampling by one power
of the system size. \cite{Bienvenu2022}

These side-chaining methods have strong parallels to  embedding methods,
\cite{Knizia2012, EmbeddingRev2016, Bootstrap2016, Bootstrap2019}
in that the side chains sample some electrons well while not sampling the ``environment''
electrons at all,
but they differ in one crucial respect.
Unlike embedding methods, in which the interaction energy between the system and its
environment is approximated, these two side-chain studies carefully set up their
overall energy estimates to be unbiased, and so they do not add any approximation.
This juxtaposition begs the question: what might we be able to gain if we instead
intentionally introduced simplifications that biased the results?
Can the cost growth with system size be reduced further while maintaining reasonable
accuracy for relevant chemical quantities like bond stretching energies?
In this study, we embark on a preliminary investigation of these questions,
concluding that, yes, for some types of predictions, a combination of correlated
sampling, side-chaining-inspired fragmentation, and embedding-inspired
Hamiltonian approximations allow for VMC cost scaling to be reduced further,
in principle to the point that it can become linear or even sub-linear
in the system size.

Unlike many embedding methods that nestle a high-level
method inside a low-level method,
such as diffusion Monte Carlo within density functional theory
\cite{doblhoff2018density}
or coupled cluster within density functional theory,
\cite{lee2019projection}
the present approach is to use the same level of theory everywhere
but to approximate its sampling and long-range electrostatics
in order to improve efficiency.
Thus, one should not expect accuracy to improve relative to
the VMC approach that is being modified.
At the same time, however, one need not rely on a lower-level
method for any part of the system.
At this early stage, we make no attempt to show in which contexts
this approach may be preferrable over more traditional embedding approaches.
Instead, this study focuses on establishing
the basic feasibility of the approach both in principle and
in practice and on delineating the conditions under which it should
be possible to reach linear and sub-linear scaling.

This paper is organized as follows.
We begin in Section \ref{sec:reducing}
with a theoretical analysis of how the
cost scaling of drawing samples in VMC can
be reduced by a combination of various techniques, within which
we include a review of the fast Slater determinant math used in
Markov chains in which only a small number of electrons move.
During this analysis, we do not address the question of accuracy,
relying instead on the assumption that certain sampling
conditions can be met without doing too much damage to VMC's
energy difference predictions.
After completing that analysis, we then 
introduce in Section \ref{sec:spotlight}
an explicit approximation scheme,
spotlight sampling, which seeks to meet these conditions while
maintaining accuracy, focusing in particular on how to deal
with both erroneous Pauli exclusion effects,
and, in Section \ref{sec:long-range}, medium and
long range Coulomb interactions.
With an approximate Hamiltonian defined, 
Section \ref{sec:cost} then completes the
cost analysis by showing that its local energy evaluations
do not alter the conclusions of the initial sampling cost analysis.
We then touch on (Section \ref{sec:stability})
how this approach intersects with the more
general VMC challenge of maintaining numerical stability
during long chains of low-rank Slater matrix inverse updates
before turning to our results.
In these results (Section \ref{sec:results}), we first verify the necessity of
spotlight sampling's buffering approach,
after which we explore proof of principle calculations
that demonstrate the cost advantages and test accuracy in
a series of alcohols' O-H bond stretches and in C=C bond stretches in examples with
differing degrees of $\pi$ system delocalization.
We then end in Section \ref{sec:conclusion}
with some concluding remarks.

%% file: theory.tex
\section{Theory}

\subsection{Reducing the Cost of Sampling}
\label{sec:reducing}

In a straightforward approach to energy estimates in
VMC, one approximates the energy as an average of the local energies
over samples drawn from the wave function's probability distribution.

\begin{equation}\label{VP}
\begin{split}
    E & = \frac{\langle \Psi | \hat{H} | \Psi \rangle}{\langle \Psi | \Psi \rangle} \\
    & = \frac{\int \Psi(R) \hat{H} \Psi(R) dR }{\int \Psi(R) \Psi(R) dR} \\
    & = \int \rho(R) \frac{\hat{H}\Psi}{\Psi} dR \\
    & = \int \rho(R) E_L(R) dR \approx \frac{1}{K} \sum_m^K E_L(R_m)
\end{split}
\end{equation}
Here $E_L (R) = \frac{\hat{H} \Psi(R)}{\Psi(R)}$ is the local energy
at the electron positions $R_m$ of the $m$th sample,
and $\rho(R)=|\Psi(R)|^2/\int|\Psi(R)|^2dR$ is the
3N-dimensional probability distribution for the $N$ electrons.
In practice, $\Psi$ is often chosen
as $\Psi(R) = e^{J(R)} D(R)$, where $D(R)$ is a Slater determinant
and $e^{J(R)}$ is a Jastrow factor,\cite{Kato, Foulkes} 
while the $K$ samples are drawn from
$\rho(R)$ via a Metropolis-Hastings Markov chain
\cite{Metro, Hastings}
in which one electron is moved at a time.
\cite{Toulouse2016}

In this straightforward approach to estimating $E$,
the cost of drawing samples is driven by the
expense of evaluating the ratio between the new and
old determinant values when one electron is moved.
After such a move, one row will have changed in
the matrix $M=X C$ of molecular orbital (MO) values,
in which $X$ and $C$ are the atomic orbital value matrix
and the matrix of occupied MO coefficients, respectively.
The ratio of the determinant values
needed by Metropolis Hastings when moving the $i$th electron
is then found (via the matrix determinant lemma) by taking
the dot product of the $i$th row of the new $M$ with the
$i$th column of the inverse of the old $M$.
\begin{equation}
    \label{eqn:one-e-move}
    \frac{D(R_{\mathrm{new}})}{D(R_{\mathrm{old}})}
    = \sum_j \hspace{1.4mm} [M_{\mathrm{new}}]_{ij}
             \hspace{1.4mm} [(M_{\mathrm{old}})^{-1}]_{ji}
\end{equation}
To reduce the cost of working with $M^{-1}$, it is updated
at $O(N^2)$ cost via the Sherman-Morrison formula after each 
accepted move. \cite{CepJas1978, UmrigarSM, LeeSM}
Generating a new and independent sample for use in Eq.\ (\ref{VP})
requires $O(N)$ of these one-electron moves,
and, at least when we are attempting to estimate the total
energy $E$, the overall number of samples $K$ needs to grow
linearly with $N$ to counteract the linear growth in the variance.
\cite{Bienvenu2022}
All told, the computational effort comes out as $O(N^4)$,
which breaks down as taking $O(N)$ samples, each of which
requires $O(N)$ moves, each of which bears an $O(N^2)$ cost
for the Sherman-Morrison update.

In this study, we will explore a combination of correlated
sampling and fragmentation in order to approximate
energy differences for local chemical changes at reduced cost.
To start, let us focus first on the effort required to draw
samples, and in particular how a fragmentation of the system
can reduce this effort.
Imagine if, as in a PMC side chain, \cite{Bienvenu2022}
we only ever had to consider situations in which a small
group of O(1) electrons had moved away from their initial
positions, while all other electrons remained frozen where they started.
In that case, only O(1) of the rows of $M$ would ever change,
allowing for an efficient evaluation of a key
determinant ratio.
\begin{equation}
    \label{eqn:fast-move}
    \frac{D(R_{\mathrm{new}})}{D(R_{\mathrm{init}})}
  = \frac{\mathrm{det}(M_{\mathrm{new}})}{\mathrm{det}(M_{\mathrm{init}})}
    = \mathrm{det}( Y \hspace{0.4mm} Z ),
\end{equation}
Here $Y$ is the rectangular matrix formed by the rows of $M_{\mathrm{new}}$
that have actually changed and $Z$ is the matrix formed
from the corresponding columns of $(M_{\mathrm{init}})^{-1}$.
The ratio $D(R_{\mathrm{new}})/D(R_{\mathrm{old}})$ needed for Metropolis is just the
ratio of two evaluations of Eq.\ (\ref{eqn:fast-move}),
and so the cost of a single-electron move is thus dominated by the
cost of forming the $O(1)$ by $O(1)$ matrix $YZ$.
If each MO is localized in space and only has nonzero contributions from 
$O(1)$ atomic orbitals, $YZ$ can be formed and its determinant taken
in $O(1)$ time.
If the MOs are not localized, then forming $YZ$ will require $O(N)$ time.
Either way, this is a reduction compared to the $O(N^2)$ time per move required by
the straightforward approach above.

Although at first glance it sounds like a drastic approximation,
our approach will put us in this computationally desirable situation
by breaking the overall system into $O(N)$ local fragments
(e.g., CH$_2$ groups), after which we will approximate each
fragment's energy using samples in which only that fragment's
and $O(1)$ neighboring fragments' electrons move.
Because other electrons are never able to relax to the
wave function's stationary distribution when evaluating
a given fragment's energy, this approach will be biased,
and we will discuss below how we minimize the effects
of this bias on our overall energy estimate.
For now, we focus on the resulting computational cost.
If each fragment contains $O(1)$ atoms and we were to take
$O(1)$ samples per fragment, we could expect the
uncertainty in each fragment's energy contribution to
be independent of system size.
However, adding these independent fragment energies together
would produce an overall energy whose uncertainty
grows as $\sqrt{N}$, and so, if the goal was to estimate
that overall energy with a fixed uncertainty,
we would instead need to take $O(N)$ samples per fragment.
Thus, if all we did was apply our fragmentation scheme
and then attempt to estimate the total energy with fixed
uncertainty, the cost would be $O(N^3)$ or $O(N^2)$
depending on whether the orbitals (and Jastrow) were nonlocal or local,
which breaks down as $O(N^2)$ samples ($O(N)$ per fragment)
that each cost either $O(N)$ or $O(1)$ each.

In this study, our goal is more specific than total energies:
by also leveraging correlated sampling, we wish to predict
energy changes due to local chemical changes at an even lower cost.
To do so, we make use of Filippi and Umrigar's space warp
correlated sampling technique. \cite{space_warp}
In this approach, whose mathematical details are in the Appendix, electron positions are sampled at one molecular geometry and ``warped'' to the other geometry in a way that leaves unchanged the positions of electrons near atoms that did not move.
We would therefore anticipate that the energy difference $\Delta E^{(l)}$ and energy
difference uncertainty $\sigma^{(l)}$ for fragment $l$ will be relatively
small if the $l$th fragment's atoms do not move during the local change,
and relatively large if its atoms do move.
If we were to sort the fragments in decreasing order of $\sigma^{(l)}$ and could bound these uncertainties by $a/l$ when using some fixed and $O(1)$ fragment sample size $K_f$, where $a$ is some constant, then estimating the overall energy difference for the local change would no longer require us to take $O(N)$ samples per fragment.  Indeed, $O(1)$ samples per fragment would now suffice.
For example, with our bound, $K_f$ samples per fragment would give an overall uncertainty of
\begin{equation}
\label{eqn:one-over-l-unc}
\sigma = \sqrt{ \sum_l (\sigma^{(l)})^2 }
       \le a \sqrt{ \sum_l \frac{1}{l^2} }
       \le \frac{\pi a}{\sqrt{6}}
\end{equation}
where we have used $\sum_{l=1}^{\infty} l^{-2}=\pi^2/6$.
With this bound, we would thus need $O(1)$ samples per fragment, or $O(N)$ samples overall, bringing our overall cost down to $O(N^2)$ for nonlocal orbitals or $O(N)$ for local orbitals. 

If the falloff in energy difference uncertainty for fragments far from the change were even faster, such as dropping off as $\sigma_l\le a/l^2$ when using $K_f$ samples per fragment, then we could afford to actually decrease the number of samples used for fragments with smaller uncertainties and lower the overall cost still further.
If, instead of taking $K_f$ samples on each fragment, we instead took $K_f/l^2$ samples, then our overall uncertainty would again be bounded by the $O(1)$ value $\pi a / \sqrt{6}$.
\begin{equation}
\label{eqn:one-over-l2-unc}
\sigma = \sqrt{ \sum_l ( l \sigma^{(l)} )^2 }
       \le \sqrt{ \sum_l \frac{a^2 l^2}{l^4} }
       \le \frac{\pi a}{\sqrt{6}}
\end{equation}
As we will see in an example below, falloff at least as rapid as $a/l^2$ can be observed in practice, implying that, at least in some applications, decaying the fragment sample size as $1/l^2$ with fragment number $l$ should still achieve an $O(1)$ uncertainty in the overall energy difference.
In such a case, the total number of samples is now just $O(1)$, again thanks to the fact that $\sum_{l=1}^{\infty} l^{-2}$ is finite, meaning that the overall cost of drawing our samples should be $O(1)$ when the orbitals and Jastrow are local and $O(N)$ then they are nonlocal.
This result, if it can be achieved while retaining accuracy, would mark a dramatic reduction in scaling compared to the $O(N^4)$ cost of the straightforward approach to drawing samples in VMC.
Towards that end, we will now turn to the details of how to efficiently evaluate accurate approximations to each fragment's local energy contribution under this aggressive sampling approach.

\subsection{Spotlight Sampling}
\label{sec:spotlight}

When approximating a given fragment's contribution to an overall energy
difference, there are two key issues that must be addressed.
First, in the space near frozen electrons, the marginal probability
distribution for electrons that are moving will be erroneously
distorted due to the over-localized Pauli exclusion holes
of the frozen electrons.
Fortunately, Pauli exclusion effects decay exponentially with
distance, which suggests that the distribution of a moving electron
should be only minimally distorted so long as it is not too close
to any of the frozen electrons.
Second, the standard VMC local energy's sum of electron-electron repulsions
will be incorrect due to the frozen electrons.
Unlike the Pauli exclusion effect, this problem decays much more
slowly with distance.
To see why, imagine modeling a long alkane chain, each of whose CH$_2$
groups is essentially nonpolar.
In our approximate approach, CH$_2$ groups with electrons frozen at
random positions would instead have large dipoles, the erroneous interactions
of which with our active fragment would decay only polynomially with distance.
In order to produce accurate estimates of the energy differences we seek,
both of these issues will need to be tackled.

\begin{figure}[ht]
\centering
\includegraphics[width=0.5\linewidth]{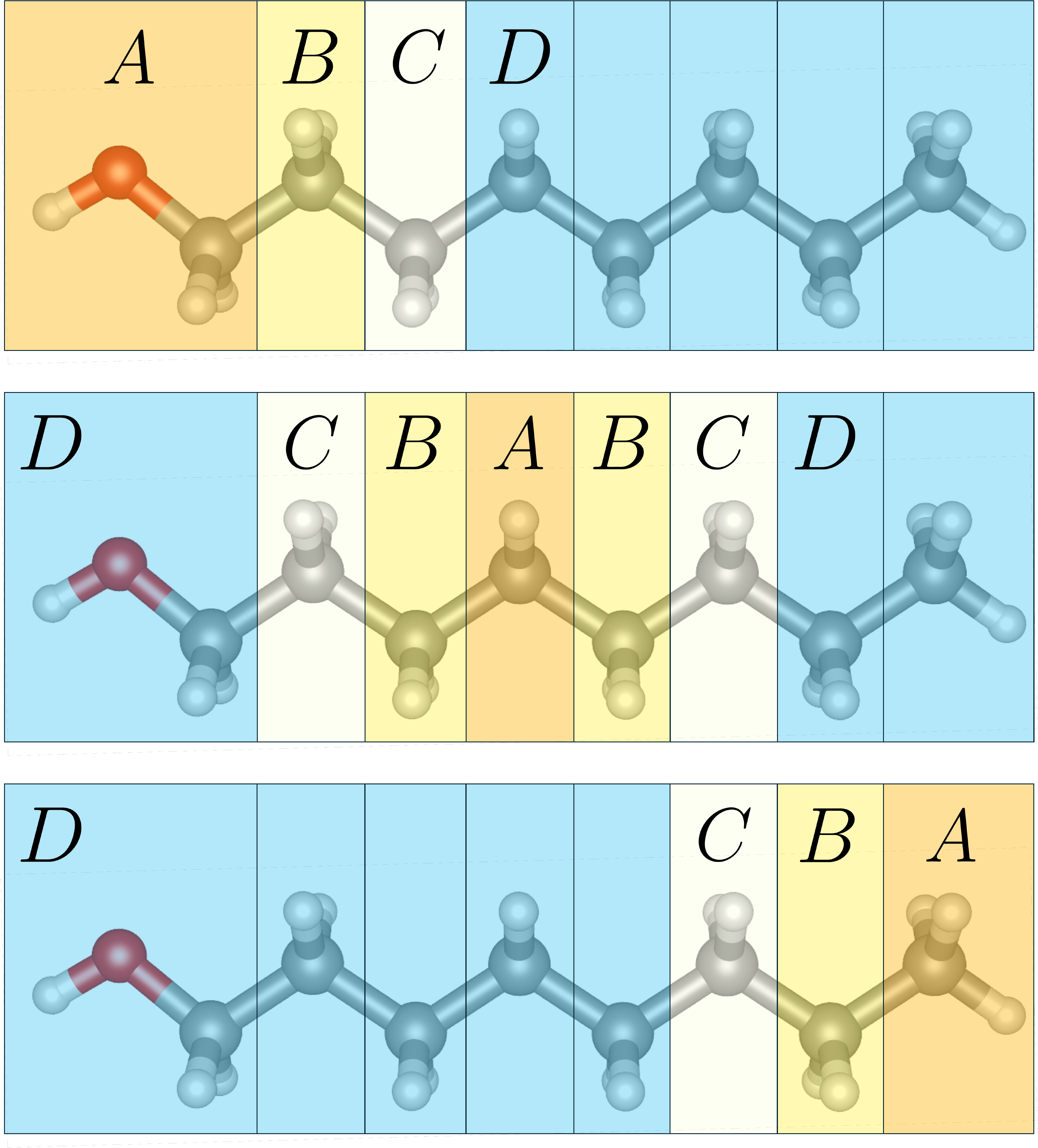}
\caption{An illustration of spotlight sampling showing three examples of
         how centering the spotlight on one fragment (region A)
         creates partially lit (region B), dimly lit (region C),
         and unlit (region D) portions of the overall system.
         The system's total energy is the sum of fragment energies, each of
         which is evaluated in its own Markov chain in which the spotlight
         centers on that fragment and in which region D electrons are frozen.
         See text for details.
    }
    \label{fig:zoning}
\end{figure}

Let's start with the Pauli exclusion problem, which we address
by adopting a ``spotlight'' sampling approach.
If we imagine shining a spotlight at the fragment whose contribution
we want to evaluate, that fragment would be brightly lit,
while fragments adjacent to it would be partially lit,
fragments adjacent to those would be dimly lit,
and the remaining fragments would be in the dark.
Grouping fragments into these four regions
--- which we will label as A, B, C, and D respectively,
as shown in Figure \ref{fig:zoning} ---
will form the foundation of our approach.
The basic idea is to hold frozen all region D electrons
while performing standard Metropolis sampling on the
electrons in regions A, B, and C.
So long as regions B and C are thick enough, the electron
positions sampled in the fragment that defines
region A should be nearly free of
the Pauli exclusion errors we are trying to mitigate.
We will therefore refer to regions B and C as our
buffer regions, and for now focus on the fact
that they protect the electrons in fragment A from
Pauli exclusion errors (we will elaborate on the division
between B and C when discussing long-range electrostatics below).

Although we now expect that an electron in region A will be mostly
free of Pauli exclusion errors, we will need a precise definition
of what it means for an electron to be ``in'' region A in order
to construct that fragment's energy difference contribution.
Towards this purpose, we define nuclear ownership factors
that smoothly partition space between the different nuclei.
Specifically, the factor $C_I(\vec{r})$ gives
the ownership fraction that nucleus $I$ has over
the point $\vec{r}$ in three-dimensional space.

\begin{equation}
\label{nuc_weights}
    C_I(\vec{r}) = \frac{e^{-d|\vec{r}-\vec{R}_I|^2}}
               {\sum_{J=1}^{N_{\mathrm{nuc}}} e^{-d|\vec{r}-\vec{R}_J|^2}}
\end{equation}
Here $d$ = 0.3 Bohr, $\vec{R}_I$ is the position of the $I$th nucleus,
and we note that, at any given position $\vec{r}$, the ownership fractions
across all $N_\mathrm{nuc}$ nuclei sum to one.
Thus, if we want to know to what degree an electron moved
to a new position is owned by region A, we can evaluate
region A's ownership fraction $w_A$ of that position as
the sum of its nuclei's ownership fractions.
\begin{equation}
    \label{eqn-A-ownership}
    w_{A}(\vec{r}) = \sum_{I\in A} C_I(\vec{r})
\end{equation}
Using this definition of ownership, we can construct an
approximate Hamiltonian operator for the fragment that defines region A.
\begin{equation}
\label{truncH}
    \hat{H}_A = \sum_{k} \frac{w_A(\vec{r}_k)}{2} \left(
                   -\xi_k \nabla^2_{k}
                   + \sum_{p\ne k} \sum_X^{ABCD} w_X(\vec{r}_p) \hspace{0.9mm}
                                          V_{AX}(\vec{r}_k, \vec{r}_p)
                \right),
\end{equation}
Here $k$ sums over both the nuclei and the unfrozen electrons,
$\xi_k$ is one for electrons and zero for nuclei,
$p$ sums over electrons, nuclei, and multipoles (see next section),
$X$ sums over $A$, $B$, $C$, and $D$,
and, finally, $V_{AX}(\vec{r}_k, \vec{r}_p)$ is the electrostatic interaction
between the particles and/or multipoles at positions $\vec{r}_k$ and $\vec{r}_p$.
It is important to note that, in the limit that region B is extended
to encompass everything other than fragment A,
pointing the spotlight at each fragment in turn and
summing up the corresponding versions of Eq.\ (\ref{truncH})
would yield the standard electronic Hamiltonian $\hat{H}$
(note that multipoles will only be present in C and D and so
can be ignored in this limit).
In other words, in the large buffer limit,
the sum of the fragment energy differences
would have the same expectation value as the
energy difference one would get from standard VMC.
The key questions we arrive at are, therefore,
(a) how large do we need to make the buffer regions in practice
to address the Pauli exclusion issue and
(b) how can we address the issue of repulsions against
frozen electrons?
We will need data to address the buffer question, so let us
focus for now on the frozen repulsions.

\subsection{Long Range Electrostatics}
\label{sec:long-range}

Naively defining $V_{AX}(\vec{r}_k,\vec{r}_p)$
as the standard two-particle Coulomb interaction will create serious errors in Eq.\ (\ref{truncH})
due to the terms in which $p$ is a frozen electron in region D
whose position is stuck at its randomly drawn initial position.
Similar errors will also occur for
electrons in region C, as their distributions will be artificially
skewed by the too-localized Pauli exclusion effects of the nearby
frozen electrons.
To address these errors, we remove the region C and region D
electrons and nuclei
from the sum over $p$ in Eq.\ (\ref{truncH}) and replace
them with multipole expansions, placing one expansion
at the nuclear center of charge
of each fragment in regions C and D.
We then define $V_{AA}$ and $V_{AB}$ as
the standard two-particle Coulomb interaction,
while $V_{AC}$ and $V_{AD}$ are defined as
the interaction energy of particle $k$ with the
multipole expansion centered at $\vec{r}_p$.
This way, we retain standard VMC interactions
between particles whose distributions we trust to be
sufficiently buffered by the presence of region C,
while resorting to multipole approximations for the
remaining interactions.
We note that, in the large buffer limit,
this approximation also gets better and better, both because
fewer interactions are replaced by multipoles and because
the multipole approximations themselves improve with distance.

To construct our multipole expansions, we perform one
round of spotlight sampling before we take the samples
that will be used for the energy difference evaluation.
During this round, when it is fragment $Y$'s
turn in the spotlight, we construct
a point-charge distribution $\phi_Y$ for that fragment
by aggregating its nuclei and the ownership-weighted
sampled positions of the unfrozen electrons.
\begin{equation}
\label{eqn:point-charge-dist}    
\phi_Y(\vec{r}) = \sum_{I\in Y} Z_I \delta(\vec{r}-\vec{R}_I)
 \hspace{0.5mm} - \hspace{0.5mm}
 \sum_{a}^{L_Y} \sum_{k}^{N_{\mathrm{u}}}
 \frac{w_Y(\vec{r}_k^{\hspace{0.4mm}(a)})}{L_Y}
 \delta(\vec{r}-\vec{r}_k^{\hspace{0.4mm}(a)})
\end{equation}
Here $Z_I$ is the charge on nucleus $I$, $L_Y$ is the number of samples
used for fragment $Y$, and $\vec{r}_k^{\hspace{0.4mm}(a)}$ is the position of the
$k$th unfrozen electron in the $a$th sample.
We then evaluate the finite multipole expansion of $\phi_X$
(we truncate after the octupole) for use in Eq.\ (\ref{truncH}).

\subsection{Overall Cost Scaling}
\label{sec:cost}

With both the Pauli exclusion errors and electron-electron repulsion
errors that arise from freezing electrons now addressed, let us
consider the cost of evaluating our energy difference.
At each individual sample of the unfrozen electron positions,
the local energy $\frac{\hat{H}_A\Psi}{\Psi}$ contains
$O(1)$ kinetic energy terms.
Each costs $O(1)$ time to evaluate if both the orbitals are local
and the locality of the Jastrow factor is exploited.
Otherwise each costs $O(N)$ time (see Appendix). 
Summing the $V_{AX}$ interactions requires $O(1)$ time
for the standard two-particle Coulomb interactions
(note that, for a given spotlight position,
$w_A$ and $w_B$ are only nonzero for $O(1)$ nuclei)
and $O(N)$ time for the charge-multipole interactions,
although this could be reduced via the
fast multipole method in sufficiently large systems.
\cite{FMM1,FMM2}
Thus, if as discussed above the uncertainty contributions
from the ordered fragments decay rapidly enough that the total
number of samples required across all spotlight positions
is $O(1)$, then the overall cost of evaluating the energy difference
for our local chemical change will add up to
$O(N)$ when nonlocal orbitals are used, and it could become
sublinear in large systems if local orbitals, Jastrow locality,
and the fast multipole method are employed.
However, the cost of first evaluating the multipoles
would still be $O(N)$, as we can expect to need $O(1)$
samples for each of the $O(N)$ multipoles.

\subsection{Numerical Stability}
\label{sec:stability}

In this spotlight approach, as in regular VMC, care is required to
mitigate the buildup of finite precision arithmetic errors when repeatedly
executing low-rank matrix inverse updates via Sherman-Morrison.
In fact, there are three contributing factors that make this
issue particularly important in the present approach.
First, the inverse matrix $(YZ)^{-1}$ on which we would like to perform
rank one updates after each one-electron move already contains a matrix inverse
inside of it: $Z$ is built from columns of $(M_{\mathrm{init}})^{-1}$.
Although we do not update $Z$ as the unfrozen electrons move around,
we do update $Y$, and in numerical testing we find that the condition
number of $YZ$ can become more extreme than that of the standard $M$
that would be used in regular VMC, leading to faster round-off buildup.
Second, in this study at least, we are employing explicitly cusped
core orbitals \cite{quady2024} whose values can range by multiple
orders of magnitude depending on how close to the nucleus an electron is.
All else being equal, a larger range of element magnitudes tends to
worsen matrix condition numbers (consider, for example, what happens
in the case of a diagonal matrix).
Third, the space warp transformation we employ in correlated sampling
can place the warped electrons near a node even when, pre-warp, the
guiding Markov chain is not close to a node.
The larger the number of electrons, the more nodes there are to
contend with, and we have observed that poor conditioning of the
post-warp $YZ$ matrix is more common in larger systems.

We address the first two issues by limiting the number of steps we
allow the Markov chain to take before explicitly re-inverting $YZ$
and the third issue by imposing an approximate modification of
our probability distribution.
Note that, we could, if we wanted, explicitly re-invert $YZ$
at every step while still maintaining the same cost scaling, as $YZ$'s
matrix dimension is $O(1)$.
However, especially when employing buffer regions, the number of
unfrozen electrons is large enough to make a Sherman-Morrison
approach worthwhile.
The question is where to strike the balance between practical efficiency
and numerical stability, and, at least in the systems tested here,
we find that re-inverting $YZ$ after each unfrozen electron has
attempted $100$ moves to be an effective compromise.
To address our third issue, we note that the conditioning of
the pre-warp and post-warp matrices is likely to differ the most
when the numerator of the correlated sampling weight factor
$W_i$ (see Appendix) is significantly different than one.
This numerator is connected to the relative condition
number, because it contains the ratio of the post-warp to pre-warp
determinants inside it.
To avoid samples that are likely to be ill conditioned,
we simply set the probability distribution equal to zero when
this numerator is less than $10^{-5}$ or greater than $10$.
This is of course an approximation, but we note that it leads to
increasing the Metropolis rejection rate by just 0.01\%,
and, as we will see in our results, it does not prevent the method
from predicting correct energy differences.
It does, however, make for a significant improvement in the
round-off error of our rank one updates, as desired. 

In the future, it will be important to test two
ideas related to numerical stability.
First, as a system gets larger and larger, more and more of the
electrons are far away from the local chemical change and thus
are unaffected by the space warp transformation.
It may be that the amount of additional difficulty created by
the corresponding correlated sampling ratios saturates in
large systems, as the positions of the far away electrons
should have less and less effect on the probability distribution
near the chemical change (this should be true for both the
pre-warp and post-warp distribution).
If the space warp effect does not saturate with increasing system
size, then an alternative correlated sampling approach may be
required in large systems.
Second, as in most areas of VMC, we expect that using
pseudopotentials will improve numerical conditioning.
In particular, they will eliminate the large orbital-value fluctuations
that come from cusped core orbitals,
which in turn should improve the conditioning of both
the Slater matrix and our $Y Z$ matrix product.
On this front, the question would be how big the effect is.
Does hiding the core electrons improve numerical stability
of spotlight sampling only a little, or does it offer large improvements?
Although our current pilot code is unable to answer these questions
at present, we flag them as important considerations for future work.

\subsection{Initialization}
\label{sec:initialization}

We begin by placing electrons according to a Lewis structure so
that the molecule is roughly spatially and spin neutral.
The position of each electron (except for those in the carbon and oxygen cores)
is then randomized within a 1 Bohr box centered on its Lewis placement. After separating the nuclei into fragments ---
in our tests, each carbon and any hydrogens, oxygens, or O-H groups
bonded to it form a fragment, as do individual H$_2$ molecules ---
we then group the initial electron positions by fragment by assigning each electron to the
closest fragment as determined by the distance to the fragment's
nuclear centroid.
We burn in the electron positions using a ``round-robin" approach in which we
loop over each fragment and, for each one, unfreeze and sample its electrons
and those of its two nearest neighbors in each direction.
At the beginning of each fragment's burn in, and once after they have all
been burned in, we re-invert $M$, for a total of $n_\mathrm{fragments}+1$
large matrix inversions.
In the systems we test, these inversions together cost much less than the initial
Hartree-Fock calculation.
After the round-robin burn in, we reassign electrons to fragments, again based on their proximities to the fragments' nuclear centroids. At this point, we have our initial positions from which we separately
launch the Markov chains that evaluate each fragment's multipole
and then continue on to evaluate each fragment's energy using
the approximate Hamiltonian in Eq.\ (\ref{truncH}).

%% file: results.tex
\section{Results}
\label{sec:results}

\subsection{Computational Details}
\label{sec:comp-detail}

We performed all quantum chemistry calculations for this project using the PYSCF package.\cite{sunLibcint2015, Pyscf2018, PySCF2020} We optimized all geometries at the MP2/6-31G level of theory unless otherwise specified. For our Hartree-Fock references, we used restricted Slater determinants with localized Foster-Boys orbitals\cite{FB} using the 6-31G(d) basis set. To ensure electron-nuclear cusps\cite{Kato} were satisfied for the VMC sampling, we employed a recently developed cusping algorithm\cite{quady2024} in which all AOs are explicitly cusped at each nucleus. We used a simple two-body Jastrow factor\cite{Foulkes, CepJas1978, CepJas1980} to satisfy the electron-electron cusps (A parameter = 0.03). Regular and spotlight sampling VMC calculations were performed using our own pilot code.

\begin{figure}[h!]
    \centering
    \includegraphics[width=0.5\linewidth]{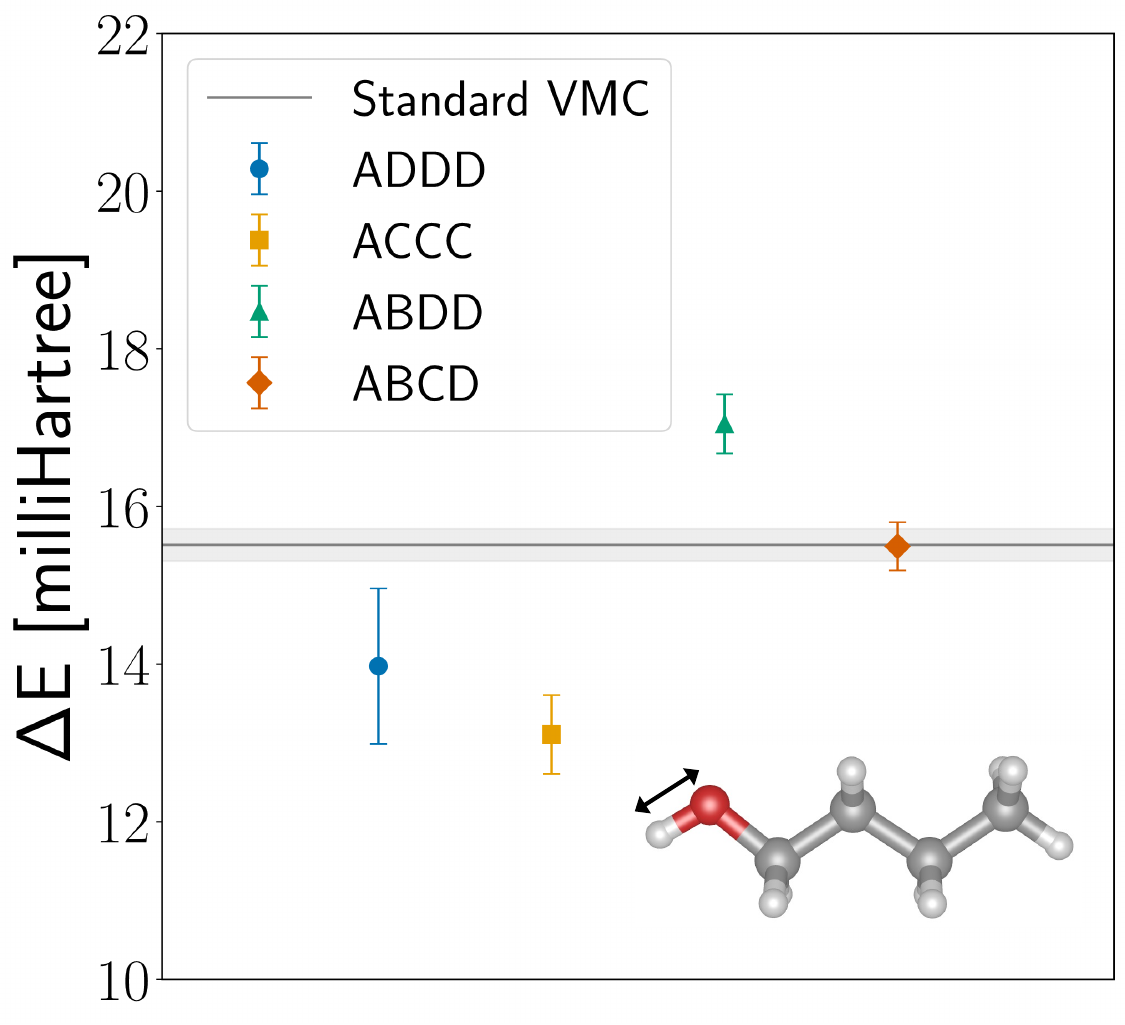}  %
    \caption{
    Predictions for the energy required to stretch an O-H bond by 0.2 Bohr
    from standard VMC and from spotlight sampling with different
    buffer region choices.
    }
    \label{fig:zoning_comp_but}
\end{figure}

\subsection{Impact of the Buffer Region}
\label{sec:buffer}

We begin by investigating the effect that different buffer region choices
have on predicting the energy difference associated with an O-H bond stretch,
as shown in
Figure \ref{fig:zoning_comp_but}.
We adopt a 4-letter naming convention for the different buffer choices we
test, in which the first letter is always ``A'' to specify that it refers
to the active fragment.
The second letter refers to that fragment's nearest neighbors, the third
to next-nearest neighbors, and the fourth to all remaining fragments.
Thus, ``ADDD'' corresponds to using no buffer region at all, with only
one fragment's electrons unfrozen in each Markov chain.
``ABDD'' also unfreezes nearest-neighbor fragment's electrons and
uses that fragment's particles for explicit Coulomb interactions, while
using multipole for next-nearest and farther fragments.
``ACCC'' unfreezes all the electrons, but employs the multipole
approximation for all non-active fragments.
Finally, ``ABCD'' unfreezes electrons in nearest and next-nearest
neighbor fragments, uses explicit Coulomb interactions for the
nearest neighbor fragment, and uses multipole for
next-nearest and farther fragments.

\begin{figure}[b!]
    \centering
    \includegraphics[width=0.5\linewidth]{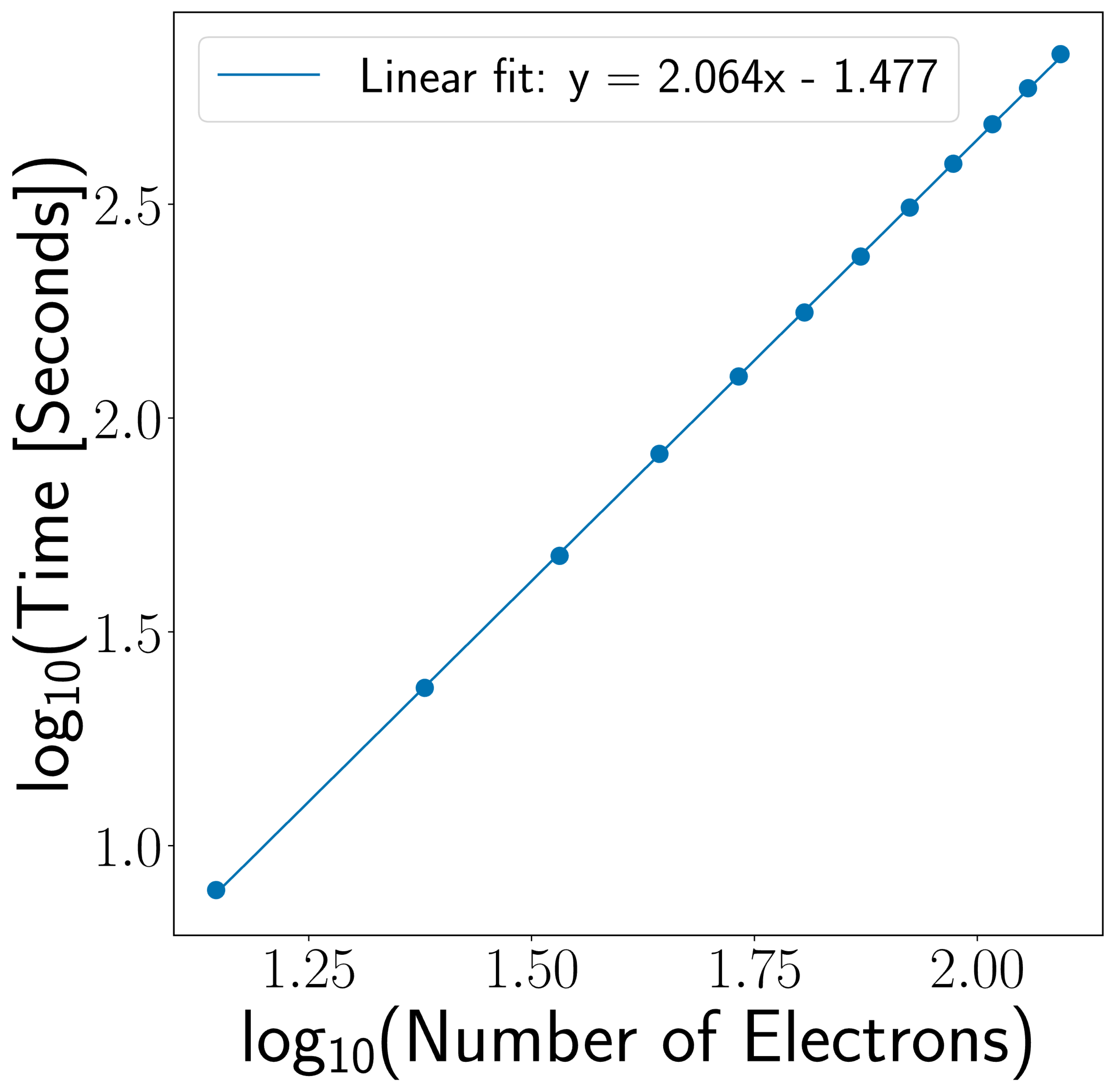}
    \caption{
    Overall spotlight sampling cost for $\mathrm{(H_2)_n}$ when employing a
    fixed number of samples per fragment.
    Each calculation used 1 core on an Intel Xeon Gold 6230 processor. 
    }
    \label{fig:timing}
\end{figure}

As seen in the figure, only the ABCD approach reproduces the predictions
of standard VMC within statistical error.
Freezing all other fragments' electrons (ADDD) produces both a significant
error and a significantly larger uncertainty, which we suspect is driven by the
effect of the randomly frozen Pauli exclusion holes in the nearest-neighbor
fragments.
The uncertainty is smaller for ACCC, in which all electrons move,
but its prediction is incorrect due to errors associated with
using a multipole approximation at short, nearest-neighbor range.
Although such errors should be reduced in ABDD, in which the
closest multipole use is next-nearest neighbor, the Coulomb interactions
with the nearest neighbor are still inaccurate due to the bias
in the nearest neighbor fragment's electron positions coming from
the frozen Pauli exclusion holes in the next-nearest neighbor fragment.
It is only when we adopt the ABCD strategy depicted in Figure \ref{fig:zoning}
that we see the spotlight sampling prediction match the standard VMC prediction.
In this approach, we mitigate multipole errors by
keeping the closest multipole use at next-nearest neighbor range
and explicit Coulomb errors by using next-nearest neighbors
as buffers in between the nearest neighbors and the frozen electrons.
We adopt this approach for all remaining calculations.

\begin{figure}[b!]
    \centering
    \includegraphics[width=0.5\linewidth]{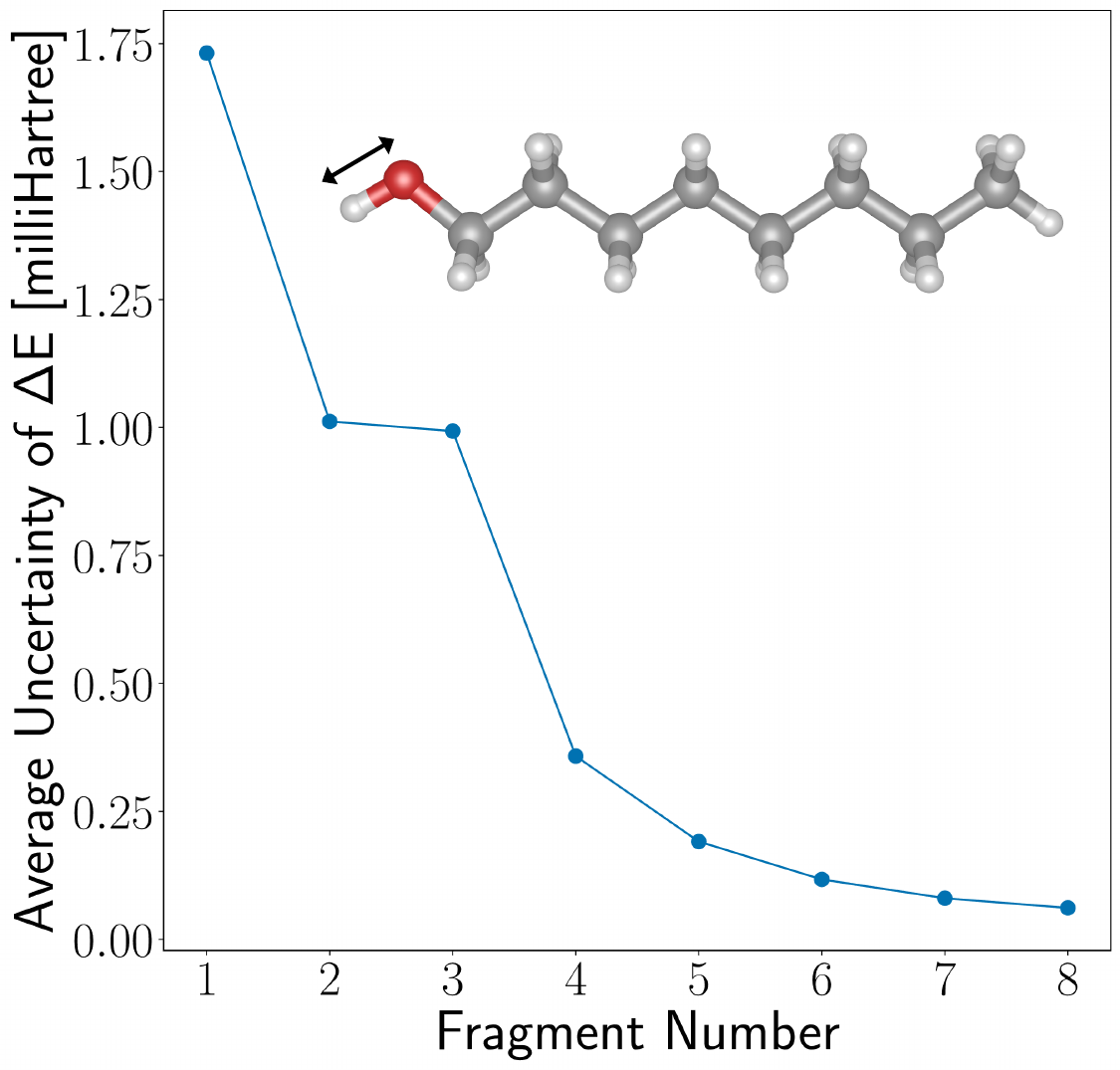}
    \caption{
    Each fragment's average uncertainty contribution to the O-H bond stretching
    energy in 1-octanol.  Each fragment contains one C atom and any H or O-H bonded to it.
    }
    \label{fig:oct_decay}
\end{figure}

\subsection{Cost Scaling}
\label{sec:results-cost}

In investigating whether observed computational cost scaling matches
the expectations from our analysis above, let us begin with a simple
test in which we take a fixed number of samples for each fragment.
As our implementation does not exploit orbital locality, we expect
each sample to carry an $O(N)$ cost, which, when combined with the
fact that there are $O(N)$ fragments, leads us to expect $O(N^2)$
cost growth overall.
We test this theory in a chain of hydrogen dimers
(bond length = 1.393 Bohr) arranged in a line,
in which each H$_2$ molecule's center is placed 4.03 Bohr
away from the previous center, and all bond axes are arranged at an
angle of 7.12 degrees relative to the line passing
through the centers.
Using each H$_2$ as its own fragment and taking 110
samples per fragment, we see in Figure \ref{fig:timing} that the
spotlight sampling cost does indeed grow as $O(N^2)$.
While our current code cannot test the reduction to $O(N)$ that would
result from local orbitals, that reduction is well established,
\cite{Bienvenu2022, Feldt2021}
and so we instead turn to the question of how effectively costs
can be reduced by taking fewer samples in fragments farther away
from the local chemical change in question.

\begin{figure}[t!]
    \centering
    \includegraphics[width=0.5\linewidth]{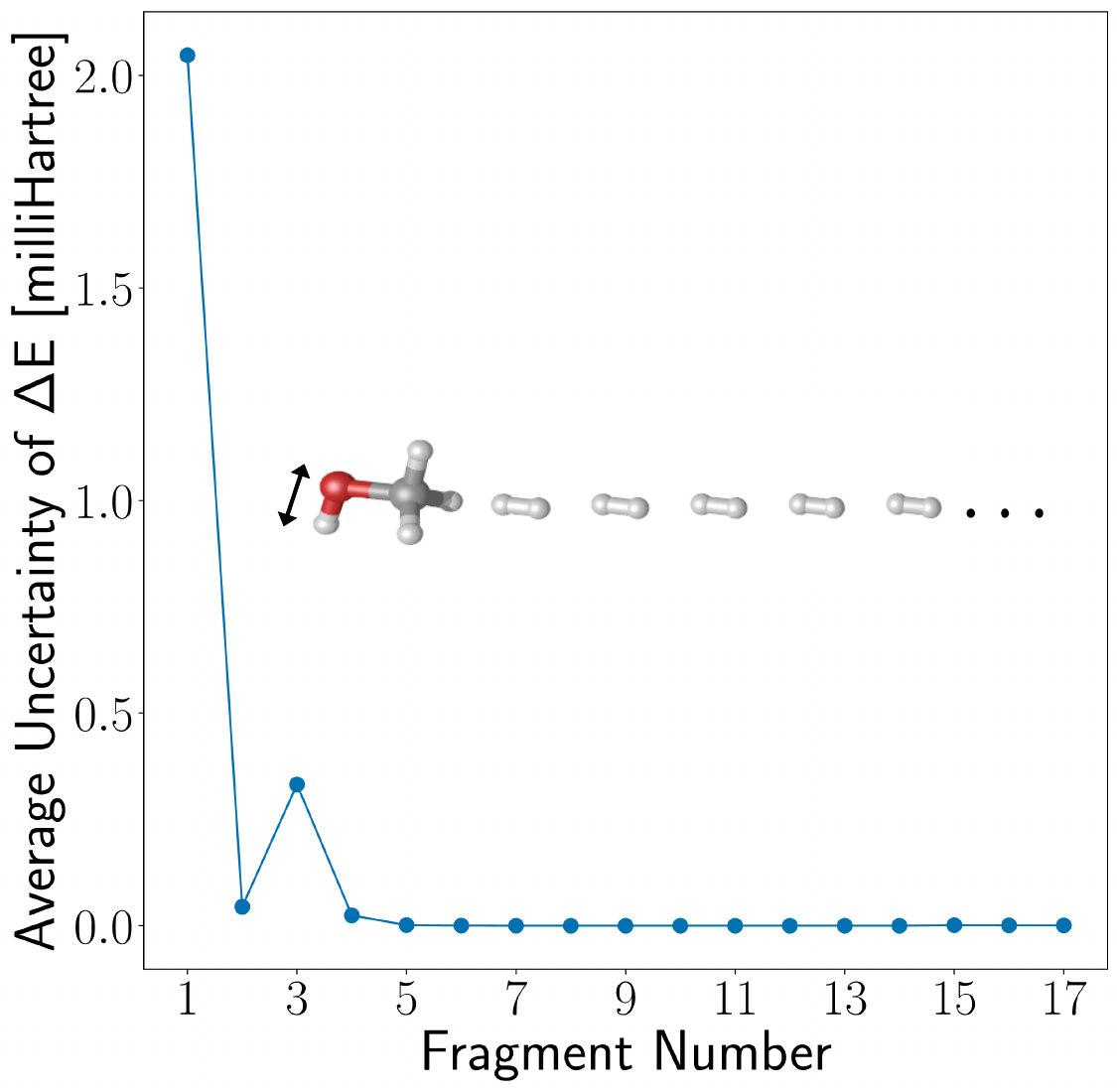}
    \caption{
    Each fragment's average uncertainty contribution to the O-H bond stretching
    energy in methanol-$\mathrm{(H_2)_{16}}$.
    Each molecule is one fragment.
    }
    \label{fig:MeOH_H2_decay}
\end{figure}

To address this question, we first perform tests of how much uncertainty each fragment
contributes to a bond stretching energy when we take a fixed number of samples per fragment.
We perform our first test in 1-octanol, where we perform 150 runs on each fragment,
in which each run uses 720,000 samples.
For each run in each fragment, we use a blocking analysis to evaluate the uncertainty
of that fragment's contribution to the bond stretching energy in that run.
The averages of these uncertainties over all 150 runs are plotted in
Figure \ref{fig:oct_decay}, where we see that the uncertainties
decrease as one moves away from the CH$_2$OH fragment
(fragment 1) that contains the bond stretch.
In fact, these uncertainties decay fast enough that they are bounded by
an $a/l^2$ function, with $a=0.01$ Hartree and $l$ the fragment number.
As discussed in Section \ref{sec:reducing}, this is a highly desirable bound,
as it suggests that decreasing the sample size for farther fragments
should be highly effective.

\begin{figure}[b!]
    \centering
    \includegraphics[width=0.5\linewidth]{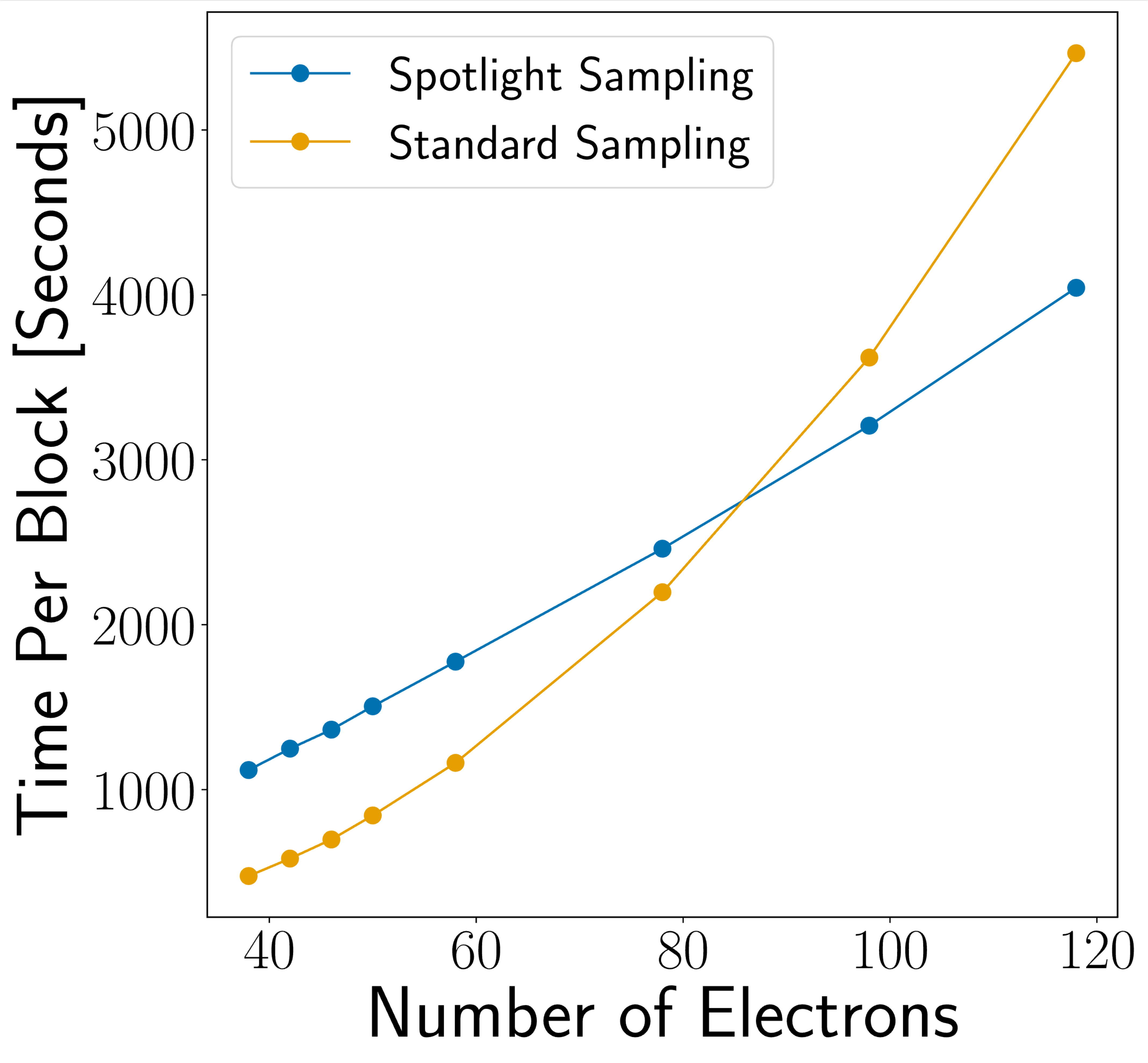}
    \caption{
Wall time comparison of spotlight versus standard correlated sampling for the O-H bond stretch in MeOH--$\mathrm{(H_2)_n}$ when employing reduced spotlight sampling efforts for farther fragments.
In all cases, 15 sampling blocks were sufficient for spotlight sampling to reach the
0.0005 Hartree uncertainty threshold, whereas the number needed
by standard sampling varied between 15 and 27.
Each calculation used five Intel Xeon Gold 6230 processors. 
    }
    \label{fig:van_vs_spot}
\end{figure}

We repeat this per-fragment uncertainty quantification in a
system in which a methanol's O-H bond is stretched when
the methanol is placed next to a line of 16 H$_2$ molecules, this time taking
480,000 samples per fragment.
In Figure $\ref{fig:MeOH_H2_decay}$, we see that the uncertainty contributions
again become small for fragments far away from the bond stretch, and indeed
we find that they can also be bounded by $a/l^2$, this time with $a=0.004$ Hartree.
Note that, in both cases, fragment 3 has a larger uncertainty contribution than
one might at first expect, but this result makes sense once we note
that the fragment 3 Markov chain contains the largest number of unfrozen electrons
and that these electrons include those of the stretched bond.

Seeing that uncertainties decay rapidly as one gets farther from the
local chemical change, we now exploit this behavior to demonstrate that,
even without making use of local orbital accelerations, spotlight sampling
can achieve lower costs than traditional VMC.
To do so, we again calculate the O-H bond stretching energy for 
methanol-$\mathrm{(H_2)_{n}}$, but this time we employ blocks of 960,000
samples for the first three fragments and much smaller blocks of 4,000 samples
for the remaining fragments, versus whole-system blocks of 960,000
samples for standard VMC.
In each system, we compute enough sampling blocks so that the overall uncertainty
in the stretching energy is below 0.0005 Hartree, always using at least 15
blocks to ensure a reliable uncertainty estimate.

In Figure \ref{fig:van_vs_spot}, we see that this test of spotlight sampling
achieves near-linear scaling with system size, with spotlight's
cost lower than standard correlated sampling once there are about 100 electrons.
The linear scaling is expected, as the vast majority of the samples are
in the first three fragments, making the total number of samples nearly
independent of system size.
Indeed, the linear cost growth comes instead from our code's $O(N)$ cost per sample.
The small quadratic term in the cost growth that begins to be visible in
the spotlight timing data in the largest system is due to the fact that
we used fixed-size 4,000 sample blocks even for the farthest fragments,
and could be eliminated by tapering this sample size further.
It is important to re-emphasize that this crossover with standard sampling
has been achieved \textit{without} exploiting local orbitals.
Were one to do so, the crossover would occur even earlier, and the cost
growth of spotlight sampling could drop to sub-linear, assuming local
Jastrows and fast multipole methods were also employed.
Although each spotlight prediction agreed with standard sampling within
error bars, this system is somewhat artificial, and so we turn to
a series of alcohols and systems with $\pi$ conjugation as more realistic tests of the method's accuracy in Section \ref{sec:accuracy}.

\subsection{Statistical Efficiency Comparison}

\begin{figure}[h!]
    \centering
    \includegraphics[width=0.5\linewidth]{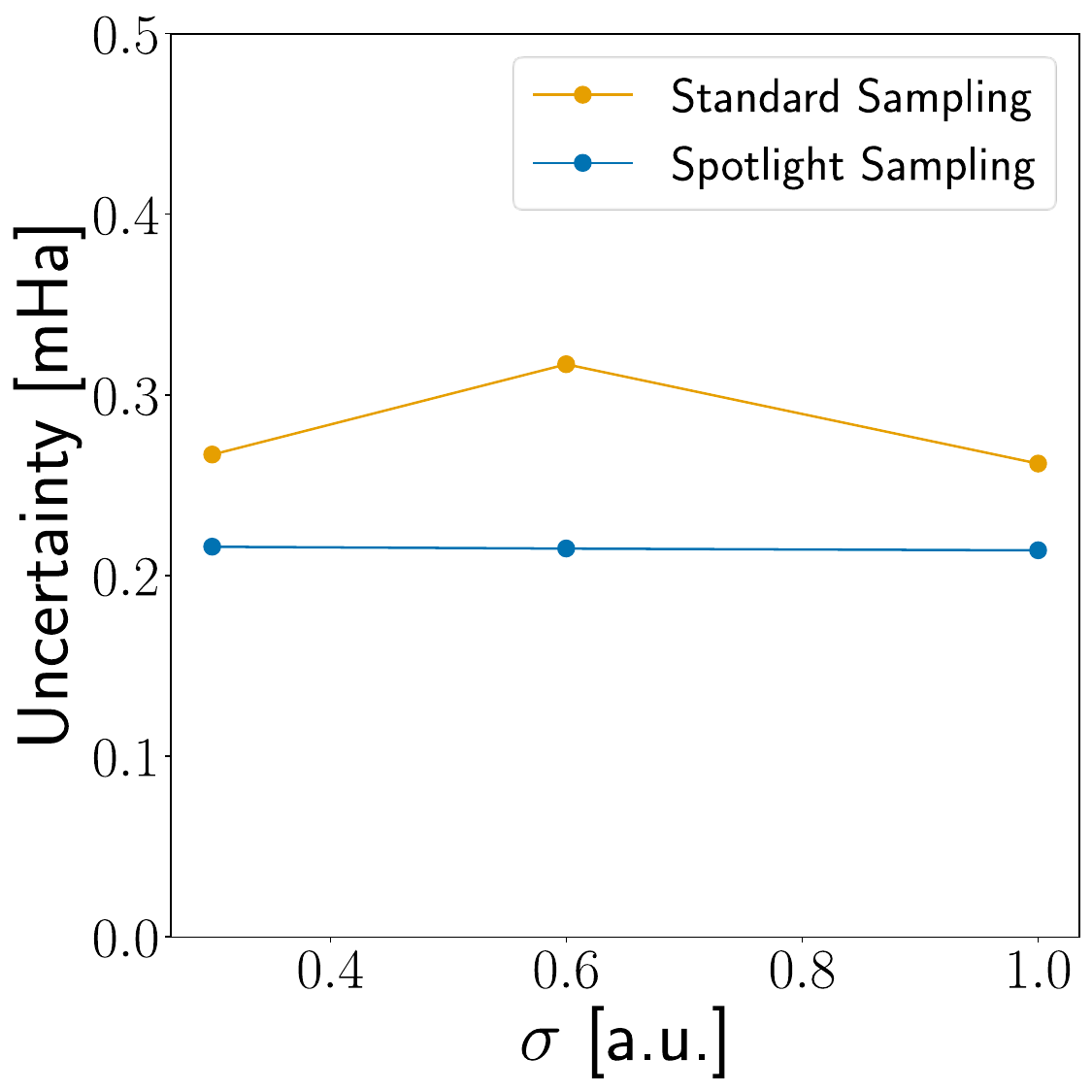}
    \caption{
    The uncertainty of the O-H bond stretch energy in 1-octanol as a function of the standard deviation of the Gaussian distribution for the 1-electron move proposals.
    }
    \label{fig:unc_vs_ts}
\end{figure}

We have also tested whether the width of the
Gaussian from which one-electron moves are drawn (which is the
analogue of the diffusion Monte Carlo timestep) has an effect on statistical
efficiency, and whether the effect is different in standard
versus spotlight sampling.
Specifically, we have performed six evaluations of the
1-octanol stretching energy, three each for standard and
spotlight sampling.
Across the three tests, we varied the standard deviation
of the Gaussian from which one-electron moves were drawn,
testing values of 0.3, 0.6 and 1.0 Bohr.
As each test used the same total number of samples,
we take the uncertainty in the stretching energy as
our measure of statistical efficiency.
Figure \ref{fig:unc_vs_ts} shows that neither
standard nor spotlight sampling had its statistical
efficiency much affected by the one-electron
draw width and that the spotlight
approach is slightly more efficient by this measure.
As seen in Table \ref{table:accept_ratios},
the acceptance ratios for standard and spotlight
sampling are almost identical and display the
expected trend of decreasing as the draw width increases.
The modest difference in statistical efficiency for
the two methods is instead more likely caused by the
fact that long-range electrostatics are handled by
multipole expansions in the spotlight approach, which
introduce different amounts of noise into the long-range
electron-electron repulsion energy than standard VMC's
approach of averaging repulsions between explicit
electron positions.

\begin{table}[htb]
\centering
\caption{Comparison of the 1-electron move acceptance ratio for different standard deviations of the move proposal distribution. For the spotlight sampling scheme, the acceptance ratio is taken as the average ratio over all fragments. 
} \label{table:accept_ratios} 
\resizebox{0.5\columnwidth}{!}{%
    \begin{tabular}{@{\extracolsep{3pt}}lcccc@{}}
        \hline \hline\\[-0.8em]
        Standard deviation [a.u.]: &  0.3 & 0.6 & 1.0 \\[0.8ex]
        \hline \\ [-1.ex]
        Accept ratio (standard)  & 0.61  & 0.42  & 0.27   \\[0.8ex]
        Accept ratio (spotlight) & 0.61  & 0.42  & 0.28   \\[0.8ex]
        \hline \hline
    \end{tabular}
}
\end{table}

\subsection{Accuracy}
\label{sec:accuracy}

\subsubsection{Alcohol Chains}
\begin{figure}[h!]
    \centering
    \includegraphics[width=0.5\linewidth]{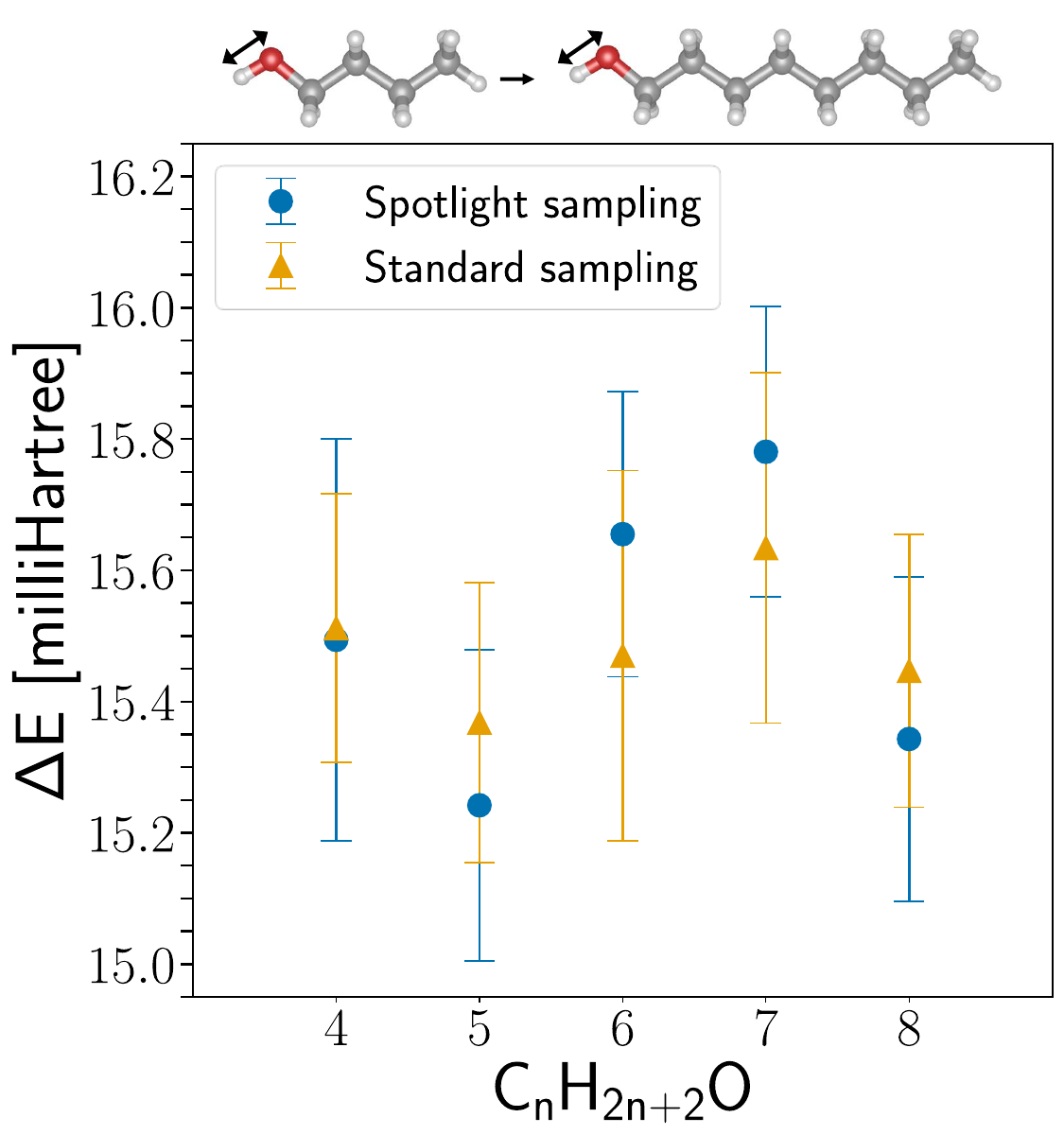}
    \caption{
    Predictions for the energy needed to stretch an O-H bond in
    the alcohols 1-butanol through 1-octanol. 
    }
    \label{fig:but_bis_oct}
\end{figure}

To test whether our approximations to the Hamiltonian and probability distribution
produce meaningful errors in predicted energy differences, we apply the spotlight
sampling approach to predict the energetic cost of a 0.2 Bohr O-H bond stretch
in a series of alcohols: 1-butanol through 1-octanol.
As seen in Figure \ref{fig:but_bis_oct}, spotlight sampling reproduces
the predictions of standard correlated sampling in all cases.
We note that this accuracy comes despite the fact that, in octanol,
the Markov chain that estimates the contribution from the fragment
containing the stretched bond froze more than half of the electrons.
 
\subsubsection{Conjugated Molecules}

\begin{figure}[h!]
    \centering
    \includegraphics[width=0.5\linewidth]{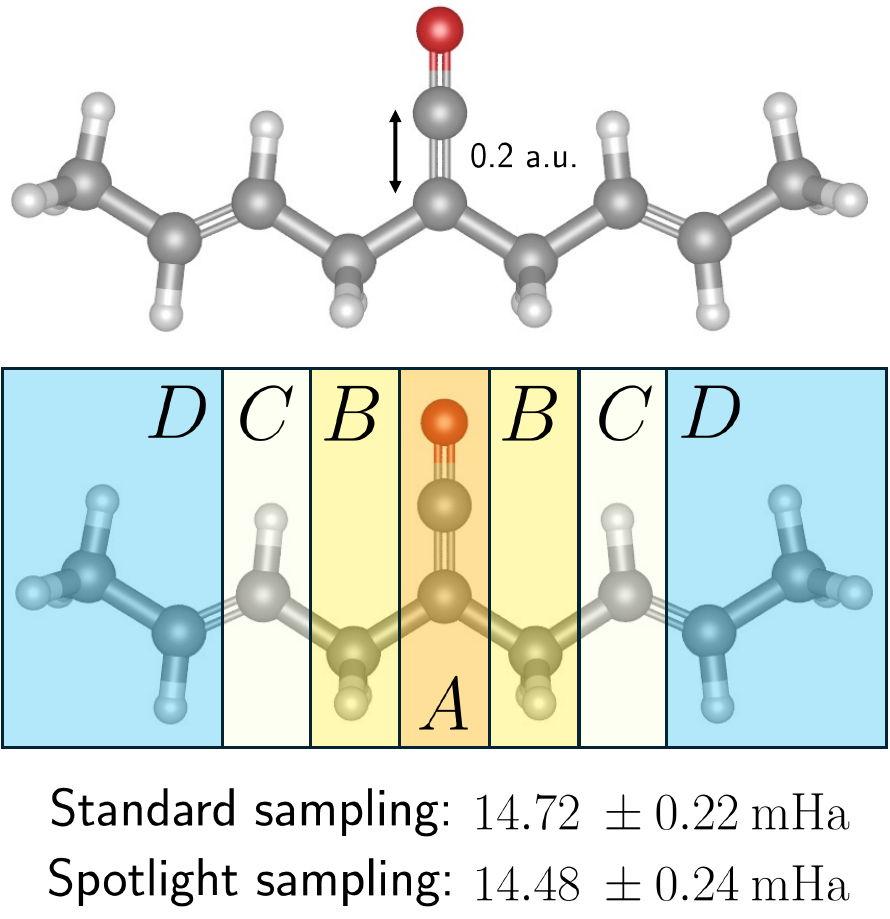}
    \caption{
    Predictions for the energy needed for a 0.2 Bohr stretch of the
    central C=C bond in CCO(CH$_2$CHCHCH$_3$)$_2$ as well as a depiction of the
    molecule and the zoning when the spotlight is placed on the central CCO group.
    }
    \label{fig:C10H14O}
\end{figure}

To test spotlight sampling's accuracy in settings where
the chemical bonding is less local in character,
we have also explored double-bond stretching in
a pair of molecules with differing degrees of $\pi$-system delocalization.
In the CCO(CH$_2$CHCHCH$_3$)$_2$ molecule shown in Figure \ref{fig:C10H14O},
the four different double bonds are separated by CH$_2$ groups into three
distinct $\pi$ systems, and the simple zoning scheme tested in this study
leads spotlight to nicely reproduce the results of standard VMC.
This success comes despite the fact that we are now stretching a double
bond and despite the fact that, when that bond is in the spotlight A
region, the other carbon-carbon double bonds are each split between
the C and D regions, meaning that some of their electrons are frozen
while others are moving.

\begin{figure}[h!]
    \centering
    \includegraphics[width=0.5\linewidth]{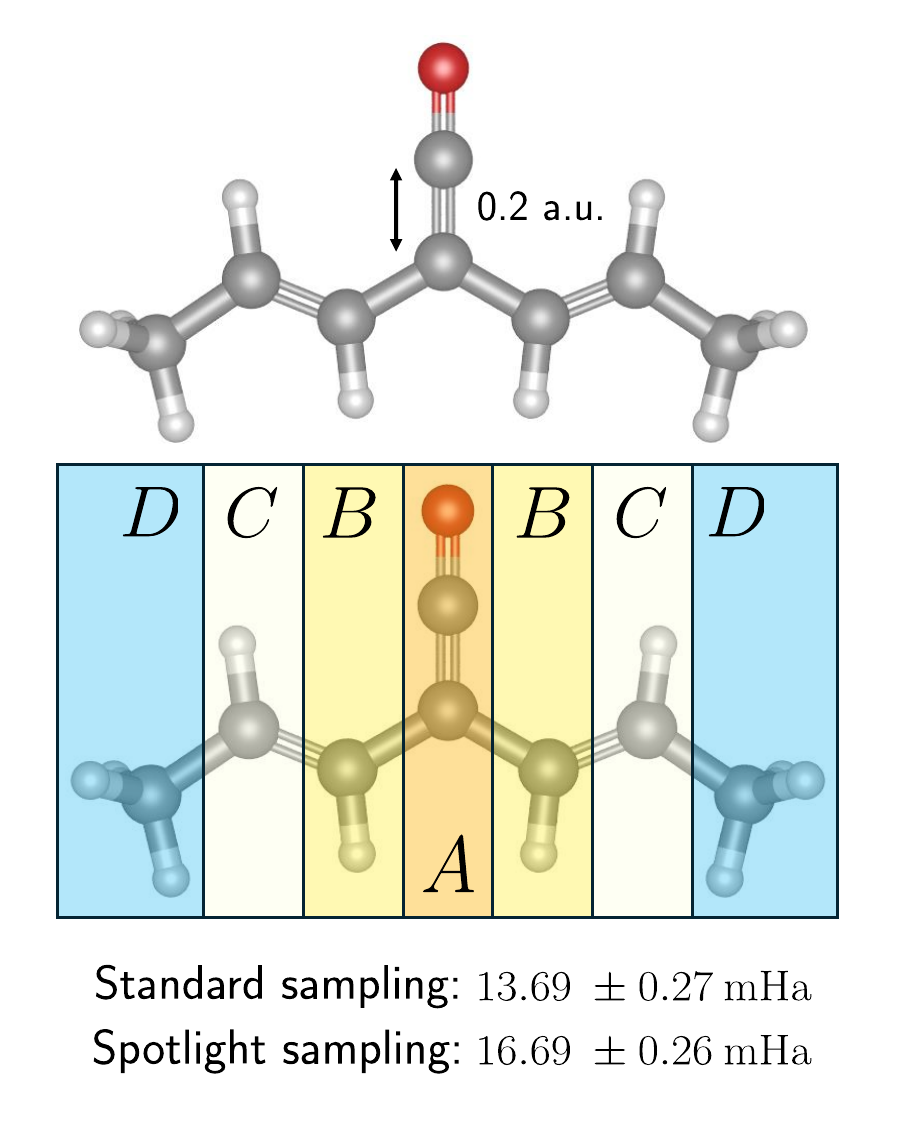}
    \caption{
Predictions for the energy needed for a 0.2 Bohr stretch of the
    central C=C bond in CCO(CHCHCH$_3$)$_2$ as well as a depiction of the
    molecule and the zoning when the spotlight is placed on the central CCO group.
    }
    \label{fig:C8H10O}
\end{figure}

In contrast, the CCO(CHCHCH$_3$)$_2$ molecule
shown in Figure \ref{fig:C8H10O}
has all four of its double bonds linked together in one delocalized $\pi$
system, and in this case the spotlight approach using our simple zoning scheme
is less accurate.
As the $\pi$ molecular orbitals now extend across most of the molecule,
the sampling of electrons in the stretched CCO group
when that group is in the spotlight
(the zoning for which is shown in Figure \ref{fig:C8H10O})
is now significantly affected by the fact that the electrons in the
methyl caps at either end are frozen.
Although they are far away from the CCO group in terms of distance
through space, the delocalized $\pi$ system extends to the edge of
those methyl groups and so experiences spurious
Pauli exclusion effects from the frozen electrons.
As it is one of the bonds in this $\pi$ system that is being
stretched, it is perhaps not surprising that accuracy suffers.

Seeing these two related double-bond stretches lead the spotlight
approach to one accurate and one inaccurate outcome raises questions
about how best to generalize and automate the zoning setup
in the future.
In particular, these results suggest that it may be important to
\textit{iterate} the spotlight scheme by asking the following question:
does the simple zoning approach used here give the same prediction as one
in which the key zones nearest the local chemical change are fused
together into a larger zone?
If not, further expansion of this central zone can be tested until the
prediction stops changing.
While this iteration may be unnecessary when delocalized $\pi$ systems
are not present, it would provide an automated way to discover the
extent of the nonlocal behavior, at least insofar as it affects the
spotlight sampling prediction.
In addition to iteration, it will likely also be important to have
two criteria for choosing which zones should be B zones for a given A
zone: the current one based on molecular connectivity but also a
second one based on distance.
In larger molecules that can fold over on themselves, bonding connectivity
may be a poor guide to which other zones will have the strongest effects
on the currently-spotlighted A zone.

%% file: conclusion.tex
\section{Conclusion}
\label{sec:conclusion}

This study has explored the efficacy of a spotlight sampling approach
to predicting the energy differences associated with local chemical
changes in VMC.
In this approach, standard correlated sampling is approximated by
using a series of separate Markov chains, one for each fragment
in the molecule, in which only the electrons near that fragment move.
Our theoretical analysis suggests that the cost of the approach can
be made linear with system size when employing nonlocal orbitals,
and that this cost should become sub-linear when
local orbitals, local Jastrows, and fast multipole methods are used.
Timing calculations show that the overall cost grows quadratically when
we take a fixed number of samples per spotlighted fragment, and that
this cost growth becomes essentially linear, as expected, once
we exploit the fact that per-fragment uncertainty decays rapidly
for fragments farther from the local change.
To maintain accuracy despite the biases inherent to freezing
most electrons' positions, we have developed an approximate
fragment Hamiltonian that employs multipoles for long range interactions.
Tests on a series of alcohols and molecules with $\pi$-conjugation show that this approach can reproduce
the predictions of standard sampling within statistical uncertainty.

In future work, there are important questions to
answer about numerical stability and the applicability of the approach
to more sophisticated wave functions.
It appears highly likely that simply employing pseudopotentials will
go a long way towards mitigating numeric challenges, but this expectation
needs to be established firmly with data.
Regarding other wave functions, a generalization of the current scheme
should be applicable to multi-determinant wave functions thanks to
the fact that, in practice, such wave functions are evaluated using
an extended Slater matrix that will have the same pattern of rows
that do and do not change when electrons in only one region are moving.
In contrast, for neural network and backflow wave functions in which moving one
electron can change every element of the Slater matrix, it may be
necessary to impose some ``row locality'' on the network setup in
order for spotlight sampling to be helpful.

%% file: appendix.tex
\section{Appendix}

\subsection{Correlated Sampling with the Space-Warp Coordinate Transformation}

We employ the correlated sampling \cite{Lowther1980} technique known as the space-warp coordinate transformation \cite{SWCT1} to reduce the statistical uncertainty of simple bond stretches (e.g., O-H stretch in 1-n-ol, n = butan-, pentan-, hexan-,...).
The relative energy difference for such a stretch, $E_s - E$,
is rewritten to be sampled from the unstretched $\Psi$ reference:

\begin{equation}\label{deltaE}
\begin{aligned}
    E_s - E & = \langle \Psi_s | \hat{H}_s | \Psi_s \rangle - \langle \Psi | \hat{H} | \Psi \rangle \\
    & \approx \frac{1}{N} \sum_{R_i\in |\Psi|^2}
    \Biggl( \frac{\hat{H}_s \Psi_s(R^s_i)}{\Psi_s(R^s_i)} W_i - \frac{\hat{H}\Psi(R_i)}{\Psi(R_i)} \Biggr)
\end{aligned}
\end{equation}
where 
\begin{equation}
    W_i = \frac{|\Psi_s(R^s_i)|^2/|\Psi(R_i)|^2 J(R_i)}{\frac{1}{N} \sum_j^N |\Psi_s(R^s_j)|^2/|\Psi(R_j)|^2 J(R_j) }
\end{equation}
$J(R_i)$ is the Jacobian of the space-warp coordinate transformation,
\cite{SWCT1, space_warp} and
\begin{align}
    \label{SWCT}
    r^s_i &= r_i + \sum_{A=1}^{n_{nuc}} (r^s_A - r_A) \omega_A(r_i) \\
    \omega_A(r_i) &= \frac{|r_i-r_A|^{-4}}{\sum_{B=1}^{n_{nuc}} |r_i - r_B|^{-4}}
    \qquad\quad \sum_{A=1}^{n_{nuc}} \omega_{A} (r_i) = 1
\end{align}
When sampling electron positions from the unstretched reference geometry, Eq.\ \ref{SWCT} shifts the electron positions for the stretched geometry to account for the displacement of the nucleus $R_A$. The weights $W_i$ are updated using the same techniques as the determinant
ratios discussed in the main text, and therefore this correlated sampling technique shares the same per-sample cost scaling as the spotlight algorithm would if used on only the
unstretched geometry.

\subsection{Kinetic Energy Evaluation}

As in many QMC contexts, evaluating the kinetic energy for the $i$th active
electron can be simplified via the following identity.
\begin{equation}
    \frac{\nabla_i^2\Psi}{\Psi} = \nabla_i^2\mathrm{ln}\Psi + (\nabla_i\mathrm{ln}\Psi) \cdot (\nabla_i\mathrm{ln}\Psi)
\end{equation}
Start with the $\nabla^2_i$ part and focus for now on the determinant
part of the wave function.
Summing the contributions over the active electrons, we have
\begin{align}
    \sum_i \frac{\nabla^2_i D(R_\mathrm{new})}{D(R_\mathrm{new})}
 &=  \sum_i \frac{D(R_\mathrm{init})}{D(R_\mathrm{new})}
            \frac{\nabla^2_i D(R_\mathrm{new})}{D(R_\mathrm{init})} \\
 &=  \sum_i \frac{\mathrm{det}(Y^{(i)} Z)}{\mathrm{det}(YZ)},
 \label{eqn:ke-YZ-ratio}
\end{align}
where we have used Eq.\ (\ref{eqn:fast-move}), and where $Y^{(i)}$
is the matrix formed by replacing the orbital values in the row of
$Y$ belonging to the $i$th electron with the corresponding orbital Laplacians.
As the matrices in the numerator and denominator of Eq.\ (\ref{eqn:ke-YZ-ratio})
differ by just one row, the matrix determinant lemma can be used to
evaluate the ratio in $O(N)$ time (or $O(1)$ time with local orbitals).
In particular, the matrix $(YZ)^{-1}$ that the lemma uses is already
available, as we are already keeping it updated via Sherman-Morrison
as discussed in the main text.
Similar arguments show that the $\nabla_i \mathrm{ln}D$ part has the
same cost scaling, and since there are only $O(1)$ active electrons, 
the overall cost scaling for the determinant part of the kinetic
energy for one sample in one fragment is $O(N)$ with nonlocal orbitals
and $O(1)$ with local orbitals.

The kinetic energy contribution of the Jastrow factor, $e^{J(R)}$, for electron $i$ requires calculating the terms
\begin{equation}\label{gradlogJ}
\nabla_iJ(R_{ij}) = \frac{A}{r_{ij}} 
\begin{pmatrix}
x_i - x_j \\
y_i - y_j \\
z_i - z_j
\end{pmatrix} \Bigg( \frac{1 - e^{-r_{ij}/F}}{r_{ij}^2} - \frac{e^{-r_{ij}/F}}{r_{ij}F}\Bigg)
\end{equation}
and 

\begin{align}
     \notag
     \nabla^2_iJ(R_{ij}) = \sum_{\alpha \in x,y,z} & \frac{A}{r_{ij}} 
     \Bigg( \frac{1 - e^{-r_{ij}/F}}{r_{ij}^2} - \frac{e^{-r_{ij}/F}}{r_{ij}F} \\
     & + (\alpha_i - \alpha_j)^2 \left[e^{-r_{ij}/F}\Big( \frac{3}{r_{ij}^3F} + \frac{1}{r_{ij}^2F^2}\Big) - \frac{3(1-e^{-r_{ij}/F})}{r_{ij}^4}\right] \Bigg) 
    \label{nablalogJ}
\end{align}
where $j$ is another electron. To ensure linear scaling for the kinetic energy evaluation for each fragment, we only evaluate Eqs. \ref{gradlogJ} and \ref{nablalogJ} for terms involving active electrons. We find this approximation to be accurate thanks to our buffer region and the fact that the Jastrow factor derivatives decay quickly with distance.

%% file: acknowledgement.tex
The authors thank Scott Garner, Leon Otis, Trine Quady,
and Connie Robinson for helpful discussions.
This work was supported by the Office of Science, Office of Basic Energy Sciences, the U.S. Department of Energy, Contract no. DE-AC02-05CH11231, through the Gas Phase Chemical Physics program. 
Calculations were performed using the Savio computational cluster resource provided by the Berkeley Research Computing program at the University of California, Berkeley and the Lawrencium computational cluster resource provided by the IT Division at the Lawrence Berkeley National Laboratory.